\documentstyle[12pt]{report}
\def\np{Nucl.\ Phys.\ }

\def\prl{Phys.\ Rev.\ Lett.\ }
\def\cmp{Commun.\ Math.\ Phys.\ }
\def\pr{Phys.\ Rev.\ }
\begin{document}
\setlength{\baselineskip}{.375in}
\begin{flushright}
YCTP-P5-92\\
hepth@xxx/9202001
\end{flushright}
\vspace{0.25in}
\begin{center}
\Large\bf
Fractional quantum Hall effect\\
and\\
nonabelian statistics\\
\vspace{0.75in}
\large\sc N. Read\\
\vspace{0.325in}
\normalsize\em
Departments of Applied Physics and Physics, P.O. Box 2157\\
Yale University, New Haven, CT 06520\\
\vspace{0.325in}
and\\
\vspace{0.325in}
\large\sc G. Moore\\
\vspace{0.325in}
\normalsize\em
Department of Physics, P.O. Box 6666\\
Yale University, New Haven, CT 06511\\
\vspace{0.75in}
\end{center}
\rm
It is argued that fractional quantum Hall effect wavefunctions can be
interpreted as conformal blocks of two-dimensional conformal field theory.
Fractional statistics can be extended to nonabelian statistics and
examples can be constructed from conformal field theory. The Pfaffian state is
related to the 2D Ising model and possesses fractionally charged excitations
which are predicted to obey nonabelian statistics.
\begin{verbatim}
Talk presented by N. Read at the 4th Yukawa International Symposium, Kyoto,
Japan, 28 July to 3 August, 1991.
\end{verbatim}
\newpage
\section*{\large\bf I. Introduction: Particle Statistics and Conformal Field
Theory}
\refstepcounter{chapter}

This paper is a brief overview of some aspects of the relationship between
the theories of the fractional quantum Hall effect (FQHE) \cite{qhe} and
two-dimensional conformal fields (CFT) \cite{cftreviews}, which has been
explored in more detail in
\cite{mooreread}. The present section describes general aspects of particle
statistics, especially nonabelian statistics; the second section uses the
Laughlin wavefunctions as an example of the relation with CFT; the third
section presents a case of nonabelian statistics in a FQHE system (the Pfaffian
state).

The notion of particle statistics in quantum mechanics usually refers to the
action of the permutation group $S_n$ on the wavefunction for a collection of
$n$ identical, indistinguishable particles: the wavefunction is taken to
transform as a definite representation of this group, which interchanges
particles. The usual examples are Bose statistics and Fermi statistics, which
are the trivial and alternating representations, respectively. A more
modern approach prefers to exchange particles along some definite paths in
space, and allows for the possibility that the wavefunction be not well-defined
when particles coincide, so that intersecting exchange paths must be omitted.
Topology then enters the subject, and while it turns out for spatial dimension
$d>2$ that the group of exchanges still reduces to $S_n$, for $d=2$ the group
of topologically distinct exchanges is Artin's braid group ${\cal B}_n$
\cite{wu}. The infinite group ${\cal B}_n$ can be generated by a set of
elementary exchanges $B_i$, $i=1,\ldots n-1$ which simply exchange particles
$i$, $i+1$ along a path not enclosing any other particles. The most familiar
representation of ${\cal B}_n$ is fractional statistics, where each $B_i$ acts
on the wavefunction as $e^{i\theta}$ ($\theta$ real) and so is a
one-dimensional, abelian representation; this includes Bose and Fermi
statistics as special cases.

We will use the term ``nonabelian statistics'' when particle
wavefunctions transform as a nonabelian representation of the permutation or
(especially) of the braid group, and we have introduced the term
``nonabelions'' for such particles \cite{mooreread}. In this case, not all the
representatives of the $B_i$ commute, and must be matrices acting on vector
wavefunctions. Thus, even when the positions and quantum numbers of the
particles have been specified, the wavefunction is not unique but is a member
of a vector space (a subspace of the Hilbert space). The vector space may be
finite dimensional but its dimension will grow with the number of particles.
Physical observables are invariant under exchanges, so cannot distinguish which
state in this space the system is in, just as the phase of the wavefunction
cannot be directly measured in the usual (one dimensional) case.
In the latter case, differences or changes of phase, and interference effects,
can, however, be observed, and analogously, nonabelian statistics can manifest
itself in physical effects.

For the permutation group, nonabelian statistics is an old idea known as
``parastatistics''. It appears, however, that parastatistics cannot be realized
in a {\em local} theory in $d>2$ in a nontrivial way (the literature containing
this result is reviewed in \cite{frohlich}). The issue is whether the global
degeneracy and nonlocal effects of exchanges are compatible with the natural
physical idea of locality of interactions, as realized for example in quantum
field theory governed by a local Lagrangian density.

An analogous analysis for the braid group in $d=2$ has been carried out in
\cite{frohlich}. It involves not only the matrices $B_i$ but also the notion of
``fusion rules''. A theory will usually contain several distinct types of
particles, each with its own $B$ matrices, and taking a particle  of one type
around one of another type will produce a matrix effect like that of $B^2$.
Fusion of two or more particles into a composite produces either annihilation
of the particles or else a particle of some new type with its own statistics
properties. The matrix elements for these processes, denoted $F$, describe the
fusion rules and will have to satisfy consistency relations involving $B$'s.
For example, taking a third particle around a pair before or after they fuse
should give the same result, since the third particle is far from the fusing
pair and so cannot distinguish the close pair from their composite (locality).
Frohlich {\it et al} \cite{frohlich} conclude that nonabelian statistics can be
acceptable in two dimensions. Indeed, local actions are known that produce
nonabelian properties: they are Chern-Simons terms for nonabelian gauge fields,
and the particle worldlines are represented as Wilson lines \cite{witten}.
Note that as in the abelian case, the statistics effects can be transferred
between the wavefunctions and the Hamiltonian by a singular gauge
transformation; the general discussion above was for the gauge choice where
statistics is exhibited in the wavefunction, {\it i.e.}\ no Chern-Simons-type
vector potentials in the Hamiltonian.

As a partial example of nonabelian statistics, we present the following
system. In the example, we may create particles $\sigma$, but only in even
numbers. The $\sigma$ particles have no
internal quantum numbers.
For two particles, the vector space for fixed positions $z_1, z_2$ is one
dimensional, and the wavefunction is $(z_1-z_2)^{-1/8}$, so
one might imagine that we have abelian fractional statistics $\theta/\pi=-1/8$.
However, for four particles, the vector space is two-dimensional. If we place
particles at $0$, $1$ and $\infty$ (which in this example can always be done
through a global conformal transformation), the wavefunctions are functions
only of the remaining coordinate $z$, and (choosing a basis) are
\begin{equation}
\psi_{\pm}(z)=(z(1-z))^{-1/8}\sqrt{1\pm\sqrt{1-z}}.
\end{equation}
In the $z$-plane, there are branch points at $z=0$, $1$. As $z$ is
analytically continued around $z=1$, we see that $\psi_{\pm}$ are interchanged.
This operation corresponds to two exchanges of the particle at $z$ with that at
$1$, {\it i.e.} to $B^2$. Other double exchanges involving $z$ are diagonal
in this basis. To exhibit $B$ itself we would have to perform a conformal
transformation to obtain $z\mapsto 1$, $1\mapsto z$. Thus we have a
single multibranched function whose two sheets are linearly independent
functions of $z$, and so the $\sigma$ particles behave as nonabelions.
For general number $2n$ of $\sigma$ particles the
wavefunctions form a $2^{n-1}$-dimensional space.

The example arises from a particular conformal field theory (CFT), the two
dimensional Ising model at its critical point. We next
describe some general features of CFT \cite {bpz}. In a
$1+1$-dimensional system with short range interactions (or a local lagrangian)
at a critical point, one has not only scale invariance but also conformal
invariance, an infinite-parameter group under which correlation functions
transform covariantly. Exactly at the critical point (so that corrections due
to slowly decaying irrelevant or marginal operators are omitted) the
correlation
functions of a collection of fields $\{\phi_{i_r}({\bf x}_r)\}$ can be split
in the form (taking the ``diagonal'' case for simplicity):
\begin{equation}
\left\langle\prod_{r=1}^n \phi_{i_r}({\bf x}_r)\right\rangle=
              \sum_p\left|{\cal F}_{p;\,i_1\ldots i_n}(z_1,\ldots z_n)\right|^2
\end{equation}
where $z_r=x_r+iy_r$. The conformal
block functions ${\cal F}_p$ are multibranched functions, analytic in their
arguments, except possibly when two $z$'s coincide. The variable $p$ labelling
different functions
runs over a finite set in a {\em rational} CFT. As the $z$'s are varied so
as to exchange some $\phi_i$'s, the
functions ${\cal F}_p$ are analytically continued to different sheets, but can
be expressed as $z$ independent linear combinations of the original functions
through some braiding matrices $B$ as we saw above. In this way, the
correlation function can be single valued.

Another operation that can be performed on the correlation functions is the
operator product expansion. As the arguments $z_1$, $z_2$ of two fields
approach one another, the operators merge into a linear combination of
single operators:
\begin{equation}
\phi_i(z)\phi_j(w)\sim\sum_k C^k_{ij}(z-w)\phi_k(w)
\end{equation}
as $z\rightarrow w$, where $\phi_k$ is some new field of type $k$ and
$C^k_{ij}(z-w)$ is a singular coefficient function. This operation can be used
to define some new matrices $F$, the fusion matrix that describes which fields
$k$ appear in the product of $i$ and $j$. In \cite{ms}
the consistency conditions that must be satisfied by $B$, $F$ were analyzed;
these same conditions
emerged later in the work of Frohlich {\it et al}. Thus CFT provides many
examples of nonabelian statistics, when conformal blocks are interpreted as
wavefuctions. In the following we show that this idea applies directly
to FQHE states.

\section*{\large\bf II. Example: Laughlin States}
\refstepcounter{chapter}

In this section we show that both the fractional statistics of
quasiparticles in the Laughlin states \cite{laugh} and the actual wavefunctions
are related to CFT conformal blocks. The construction given here is from
\cite{mooreread}, but parts of it have been obtained independently by others.

In $1+1$-dimensional Euclidean spacetime, define a free scalar field by its
correlator,
\begin{equation}
\left\langle\varphi(z)\varphi(z')\right\rangle=-\log(z-z')
\label{bosecorr}
\end{equation}
where the $\log$ is complex, so the field creates right-moving excitations
only. All correlators can be obtained from this one using Wick's theorem.
Now consider the function
\begin{equation}
\left\langle\prod_{i=1}^N{\rm e}^{i\sqrt{q}\varphi(z_i)}
\,{\rm e}^{-i\sqrt{q}\int d^2z'\,\bar{\rho}\varphi(z')}\right\rangle
\end{equation}
where $\bar{\rho}=1/2\pi q$, $q$ is an integer, and the integral is taken over
a disk of area $2\pi qN$ centered at the origin. We expand and contract using
(\ref{bosecorr}). Each exponential is assumed
normal ordered, {\it i.e.\/} $\varphi$'s from the same exponential are not to
be contracted together. The result is \cite{coleman}
\begin{equation}
\prod_{i<j}(z_i-z_j)^q\,{\rm e}^{-\frac{\scriptstyle1}{\scriptstyle2\pi}
\sum_i\int d^2z'\,\log(z_i-z')}.
\label{laughcorr}
\end{equation}
In the last factor, the real part of the $\log$ gives ${\rm e}^{-\frac{1}{4}
\sum_i |z_i|^2}$ for $z_i$ inside the disk. Apart from the remaining phase, we
now recognize the function as Laughlin's wavefunction for particles in the
lowest Landau level at filling factor $\nu=1/q$. Since the vertex operators
$\exp{i\sqrt{q}\varphi(z)}$ are associated with Coulomb charges, it is clear
that we have a holomorphic version of Laughlin's 2D plasma picture
\cite{laugh}, which gives the wavefunction and not just its modulus squared!

The imaginary part of the log in (\ref{laughcorr}) winds
by $2\pi d^2z'$ as each $z_i$ goes round each point $z'$ in the integration
region, so is highly singular \cite{semenoff}. It contributes a pure phase to
the function, which can be gauged away by an equally singular gauge
transformation. This phase describes a uniform  magnetic field, which we
identify as the physical magnetic field in the FQHE problem. A nice way to
see this is to consider not the phase itself but the {\em change} as one
particle describes a closed loop $C$. Clearly the exponent changes by
\begin{equation}
-\frac{1}{2\pi}\int d^2z'\,2\pi i
\end{equation}
where the integral is over the area enclosed by $C$, so the phase change is
$1/2\pi$ times the area of the loop, which is just the Berry phase for
adiabatic transport of a particle of charge $1$ in a field of strength
$1/2\pi$.
In all the following equations, the
singular phase will be implicitly gauged away.

In a similar way, one can obtain quasihole wavefunctions as (for two
quasiholes):
\begin{eqnarray}
\lefteqn{\left\langle{\rm e}^{{i\over\sqrt{q}}\varphi(z)}
{\rm e}^{{i\over\sqrt{q}}\varphi(w)}
\prod_{i=1}^N{\rm e}^{i\sqrt{q}\varphi(z_i)}
\,{\rm e}^{-i\sqrt{q}\int
d^2z'\,\bar{\rho}\varphi(z')}\right\rangle}\nonumber\\
&&=(z-w)^{1/q}\prod_{k}(z-z_k)(w-z_k)
\prod_{i<j}(z_i-z_j)^q\,{\rm e}^{-\frac{1}{4}\sum_i |z_i|^2-\frac{1}{4}|z|^2
-\frac{1}{4}|w|^2}.
\label{twoqholes}
\end{eqnarray}
The first factor gives explicitly the fractional statistics, $\theta/\pi=1/q$.
Apparently, the constructions from CFT correlators give results in the gauge
where all Berry phases---background magnetic field as well as fractional
statistics---appear in the wavefunction, not as a vector potential (connection)
in the Hamiltonian. This will be important to us in the next section when we
move on to nonabelian statistics (adiabatic transport).

We now return to the
$2+1$-dimensional world of electrons in two dimensions in a strong magnetic
field
and describe briefly the order parameter picture of the FQHE,
following \cite{nr} (see also \cite{gm,zhk}). In words, a composite of $1$
electron and $q$ quasiholes (or ``vortices'', or ``flux quanta'') at
filling factor $\nu=1/q$, ({\it i\/}) is a boson; ({\it ii\/}) sees no net
effective magnetic field. When both ({\it i\/}) and ({\it ii\/}) are true,
the composite can ``Bose condense'', {\it i.e.\/} have long-range order. For
the Laughlin states, ({\it i\/}) and ({\it ii\/}) follow from the calculations
in \cite{asw}.

More formally, let $\psi^\dagger$ create an electron in the lowest Landau
level (LLL), and let
\begin{equation}
U(z)=\prod_{i=1}^N(z-z_i)
\end{equation}
be Laughlin's quasihole operator (acting in the LLL). If $\left|0_{\rm
L};N\right\rangle$ is the normalized Laughlin state for $N$ particles, then one
can show \cite{nr}
\begin{equation}
\lim_{|z-z'|\rightarrow\infty}\lim_{N\rightarrow\infty}
\left\langle 0_{\rm L};N \right|U^\dagger(z)^q\psi(z)\psi^\dagger(z')U(z')^q
\left|0_{\rm L};N\right\rangle{\rm e}^{-\frac{1}{4}|z|^2-\frac{1}{4}|z'|^2}
=\bar{\rho}.
\end{equation}
Given our choice of order parameter operator, the Laughlin state
is the state with the most order. In fact, we have \cite{nr}
\begin{equation}
\left|0_{\rm L};N\right\rangle=\left(\int d^2z\,\psi^\dagger(z)U(z)^q{\rm e}^
{-\frac{1}{4}|z|^2}\right)^N\left|0\right\rangle
\end{equation}
where $|0\rangle$ is the vacuum (no electrons). This says that the Laughlin
state is precisely a Bose condensation of the composite bosons.

The order parameter in a general state obeys \cite{nr} a system of
Landau-Ginzburg-Chern-Simons equations \cite{girvin}, which also involve
internal vector and scalar potentials. Similar equations are found in
another, more field theoretic approach \cite{zhk,leefish} which has been
reviewed recently by Zhang \cite{zhang}. The ideas have been extended
\cite{leefish,nr2,blokwen} to the
hierarchy scheme \cite{hald,halp,varhier,jain} and its generalizations
\cite{nr2},
which give states with abelian
statistics for all filling factors $p/q<1$. These also have interpretations
as CFT correlators \cite{mooreread,nr2}.

\section*{\large\bf III. Pfaffian State and Nonabelions}
\refstepcounter{chapter}

Consider the wavefunction \cite{mooreread}
\begin{equation}
\Psi_{\rm Pf}(z_1,\cdots,z_{N})=\hbox{Pfaff}
\left({1\over z_i-z_j}\right)
\prod_{i<j}(z_i-z_j)^q \,{\rm e}^{-{1\over4}\sum |z_i|^2}
\label{pfaffstate}
\end{equation}
The Pfaffian is defined by
\begin{equation}
\hbox{Pfaff}\,M_{ij}={1\over 2^{L/2}(L/2)!}\sum_{\sigma\in S_L}{\rm
sgn}\,\sigma \>
\prod_{k=1}^{L/2}M_{\sigma(2k-1),\sigma(2k)}
\end{equation}
for an $L\times L$ antisymmetric matrix whose elements are $M_{ij}$, or as the
square root of the determinant of $M$; $S_n$ is the permutation group on $n$
objects. $\Psi_{\rm Pf}$ can be regarded as a wavefunction for
spinless or spin polarized electrons in the LLL if $q>0$ is even, since then
it is antisymmetric; the filling factor is $\nu=1/q$. Note that
$\hbox{Pfaff}\,((z_i-z_j)^{-r})$, $r$ odd, $<q$ would also give a valid
wavefunction.

The idea behind the construction of this wavefunction was the following. From
the calculations in \cite{asw}, it follows that, in an incompressible
fluid state of electrons at filling factor $1/q$, the composite
$\psi^\dagger(z)U(z)^q$ is a neutral boson if $q$ is odd (as mentioned in \S
II) and is a neutral fermion if $q$ is even. Hence, for $q$ odd the operator
can Bose condense, which gives the Laughlin states. For $q$ even, on the other
hand, it cannot condense singly, but pairs of fermions can condense, as in the
BCS theory of superconductivity. For the spinless or spin-polarized case, the
pairing function must be of odd parity to satisfy Fermi statistics. Thus a
possible function is
\begin{equation}
\left(\int d^2z d^2w \,{1\over z-w} \,\psi^\dagger(z)U^q(z)
\psi^\dagger(w)U^q(w) {\rm e}^{-{1\over 4}(|z|^2+|w|^2)}\right)^{N/2}|0\rangle.
\end{equation}
Writing out this function in co-ordinate representation yields
(\ref{pfaffstate}). The Pfaffian is precisely the sum of products of fermion
pairs, antisymmetrized over all distinct ways of pairing. More generally, the
pairing function $(z-w)^{-1}$ can be replaced by $(z-w)^{-r}$
with $r$ odd,  provided $r<q$, as noted above. This construction
was also inspired by noticing that the Haldane-Rezayi (HR) spin-singlet state
for electrons at $\nu=1/q$, $q$ even \cite{haldrez}, can be written in the form
involving $\det(z_i^\uparrow - z_j^\downarrow)^{-2}$. The real-space
form of the spin singlet BCS wavefunction for spin-${1\over 2}$ fermions
is just such a determinant of an even parity pairing function, as is
well known (it is perhaps less well known that the analogous result for
spinless or spin-polarized fermions is a Pfaffian). Thus the HR state is a
condensate of pairs of spin-${1\over 2}$
composite neutral fermions, $\psi^\dagger_\sigma(z)U(z)^q$ \cite{mooreread}!

The pairing picture suggests two kinds of possible excitations. One is the
analogue of the BCS quasiparticle, obtained by adding composite fermions, or
by ``breaking pairs''. Thus a state with such a neutral
fermion excitation localized at $z$ is obtained by acting on the ground state
with $\psi^\dagger(z)U(z)^q$. (In the HR state, one likewise obtains
spin-${1\over 2}$ fermions.) These excitations have no analogue in the Laughlin
states since there the bosons are already condensed; the only excitations,
other than the collective density mode \cite{gmp}, are Laughlin's
quasiparticles, which are fractionally charged. In the order parameter picture,
these are vortices, {\it i.e.\/} the order parameter winds in phase by a
multiple of $2\pi$ around each quasiparticle, which thus resemble flux quanta
in
a superconductor, the flux quantum being $\Phi_0=hc/e$ since the order
parameter
carries charge $1$ from the single electron that it contains. The flux $\Phi$
determines the charge $e^*$ through the quantized Hall relation,
$e^*=\nu(\Phi/\Phi_0)$. Similarly, in the paired states, one expects to find
excitations corresponding to flux quantized in multiples of ${1\over 2}\Phi_0$,
since the order parameter contains two electrons, and these will have charge in
multiples of $(2q)^{-1}$.

These ideas motivated the following wavefunction \cite{mooreread} for a pair
of quasiholes in the Pfaffian state:
\begin{eqnarray}
\lefteqn{\quad\Psi_{\rm Pfaff + qholes}(z_1,\ldots,z_{N};v_1,v_2)=}\\
&&\left\{\sum_{\sigma\in S_N}{{\rm sgn}\,\sigma\,
\prod_{k=1}^{N/2}[(z_{\sigma(2k-1)}-v_1)
(z_{\sigma(2k)}-v_2)+(v_1\leftrightarrow v_2)]
\over (v_1-v_2)^{{1\over 8}-{1\over {4q}}}(z_{\sigma(1)}-z_{\sigma(2)})
\cdots (z_{\sigma(N-1)}-z_{\sigma(N)}) }
\right\}
\prod_{i<j}(z_i-z_j)^q {\rm e}^{-{1\over 4}\sum_i |z_i|^2-{1\over 4}|v_1|^2
-{1\over 4}|v_2|^2}\nonumber
\label{pfaff+qholes}
\end{eqnarray}
The factor in brackets can be rewritten as
\begin{equation}
(v_1-v_2)^{{1\over 4q}-{1\over 8}}
\hbox{Pfaff}\,\left({(z_i-v_1)(z_j-v_2)+(v_1\leftrightarrow v_2)\over
z_i-z_j}\right).
\label{altform}
\end{equation}
(The reason for the assumed form of the exponent of $(v_1-v_2)$ will be
explained below.)
The extra factor in each term of the Pfaffian resembles a pair of Laughlin
quasihole operators, except that in each term each factor acts on only one
member of each pair of fermions. Therefore, in an average sense, each
quasihole is a half flux and carries charge $1/2q$. Note that this
construction cannot produce one quasihole. One way to see that this is
impossible is that working on a compact geometry like the sphere, the total
flux is quantized in units of $\Phi_0$. Another feature of (\ref{pfaff+qholes})
is that as $v_1\rightarrow v_2$, we recover a Laughlin quasihole of charge
$1/q$.

While our construction of the Pfaffian state was motivated by order parameter
considerations, we can also give an interpretation using conformal field
theory, which then suggests the full structure of the system of excited
quasihole states. Introduce free, massless real (Majorana) fermions $\chi$ in
$1+1$ dimensions:
\begin{equation}
\left\langle\chi(z)\chi(z')\right\rangle={1\over z-z'}
\end{equation}
Then using also the free scalar field as before, construct
\begin{equation}
\left\langle\prod_{i=1}^N \chi(z_i){\rm e}^{i\sqrt{q}\varphi(z_i)}\,
{\rm e}^{-i\sqrt{q}\int d^2z'\,\bar{\rho}\varphi(z')}\right\rangle.
\end{equation}
This reproduces the FQHE wavefunction (\ref{pfaffstate}). To reproduce the two
quasihole function (\ref{pfaff+qholes}), we should understand the CFT we are
dealing with. Majorana fermions arise naturally in the two-dimensional Ising
model at its critical point. The fermions, in conjunction with their leftmoving
partners $\bar{\chi}(\bar{z})$, represent the energy density fluctuation
$\varepsilon(z,\bar{z})=\bar{\chi}(\bar{z})\chi(z)$ which hence has dimension
$x=1$ and so the correlation length exponent is $\nu=(2-x)^{-1}=1$. The other
fields in the Ising model are the spin fields, {\it i.e.\/} the Ising spin
itself. These are operators which produce a square root branch point in the
fermi field \cite{bpz}. With the latter we can reproduce the two quasihole
state:
\begin{equation}
\Psi_{\rm Pfaff + qholes}=
\left\langle\sigma(v_1){\rm e}^{{i\over 2\sqrt{q}}\varphi(v_1)}
\sigma(v_2){\rm e}^{{i\over 2\sqrt{q}}\varphi(v_2)}
\prod_{i=1}^N \chi(z_i){\rm e}^{i\sqrt{q}\varphi(z_i)}\,
{\rm e}^{-i\sqrt{q}\int d^2z'\,\bar{\rho}\varphi(z')}\right\rangle.
\end{equation}
The square roots produced by the spin fields $\sigma$ are cancelled by those
produced by the ``half fluxes'' $\exp({i\over 2\sqrt{q}}\varphi(v_1))$, so
the electron wavefunctions are single-valued. The factor $(v_1-v_2)^{-{1\over
8}}$ in (\ref{pfaff+qholes}) is now explained; it was inserted to make the
equality with the correlator hold, and reflects the fact that the conformal
weight of the spin field is $1/16$ (and hence $\eta=1/4$ in the Ising model).

The generalization to $2n$ of our ``half flux'' quasiholes now seems
self-evident: we should insert $2n$ of the combinations
$\sigma\exp({i\over 2\sqrt{q}}\varphi)$ into the correlator. However, it is
known \cite{bpz} that the Ising model part of such an expression is ambiguous;
for fixed positions $v_1$, \ldots $v_{2n}$ of the spin fields there are many
different possible correlators (conformal blocks), forming a vector space of
dimension $2^{n-1}$ (over the complex numbers) independent of how many
fermions $\chi$ are present.
To resolve the ambiguity of notation, the spin fields should be
replaced by ``chiral vertex operators'' \cite{ms}. Furthermore, the monodromy
of these functions is nonabelian, {\it i.e.}\  analytically continuing $v_i$
around $v_j$ produces a linear combination of the original branches of the
functions, the coefficients being some braiding matrices that don't commute.
At present we do not know
the explicit functions for $2n$ spin fields and $N$ fermions in general,
except in the case $N=0$. The case $N=0$ may seem strange from the point of
view of electron wavefunctions, but it does at least provide an explicit
realization of nonabelian statistics, which is just that given in \S I, where
the wavefunctions for the ``$\sigma$ particles'' are just $N=0$ Ising conformal
blocks. In fact, the braiding properties of the spin fields are independent
of $N$.
Note that, when discussing the $N$ fermion
functions as electron wavefunctions, we must show that they are $2^{n-1}$
linearly independent functions {\em of the electron co-ordinates} but this
is easily done using the operator product expansion of the spin fields as
the co-ordinates $v_1$, \ldots $v_{2n}$ approach each other in pairs
$v_1\rightarrow v_2$, $v_3\rightarrow v_4$, etc.\ \cite{readun}, even without
full knowledge of these functions. The same operator product expansion also
shows that as any $v_i$ approaches any $v_j$, the leading term is equivalent
to a Laughlin quasihole of charge $1/q$, which is clearly a desirable property.
This completes our construction, though it remains to check that adiabatic
transport of our quasiholes does give the same nonabelian statistics as the
monodromy of the conformal blocks. We are confident that this is true because
of the existence of $2^{n-1}$ functions and because analogous properties hold
in the abelian examples of \S II.

Before closing, some comments on recent related work. Wen \cite{wen} has
argued that excitations of wavefunctions arising in one of Jain's later
constructions \cite{jain} possess nonabelion excitations, using the point of
view of ref \cite{mooreread} of wavefunctions as conformal blocks. These
examples involve SU(N) symmetry. Greiter, Wen and Wilczek \cite{gww} have
studied numerically the vicinity of $\nu=1/2$ for certain Hamiltonians in
spherical geometry, and find
evidence for an incompressible state with the excitations (neutral fermions and
charge $1/4$ quasiholes) of the Pfaffian state. However, they claim that the
quasiholes obey abelian $\theta/\pi=1/8$ statistics. Their arguments are based
on an ``adiabatic heuristic'' principle that connects the FQHE state to a
paired state of fermions in zero magnetic field given by the Pfaffian alone.
While this picture is extremely similar to the order parameter or condensation
picture put forward in previous papers \cite{nr,gm,zhk,mooreread} and herein,
we do not agree with the conclusion that the statistics is abelian. The full
structure of the Landau-Ginzburg-Chern-Simons theory of the state must account
for the fermion excitations, and not only the paired order parameter.
Their conclusion is what one would obtain by neglecting the Pfaffian completely
when calculating the statistics of the quasiholes by adiabatic transport {\it
\`{a} la} \cite{asw}.
In fact, while a
direct search for nonabelian statistics must use wavefunctions for 4 quasiholes
(which are not given by Greiter {\it et al}), there is already a difference in
the apparent abelian statistics for two quasiholes, as shown by our eq.
(\ref{altform}), which implies $\theta/\pi=1/8-1/8=0$; this effect would
be due to the Pfaffian.
These authors also point out that
the Pfaffian state is the unique highest density zero energy eigenstate of a
certain {\em three-body} Hamiltonian.

Clearly, once the possibility of nonabelions is accepted, many questions
remain. How would we confirm their statistics in an experimental setting?
Is a gas of nonabelions a superfluid, as for anyons
\cite{laughany}?

{\bf Acknowledgements}: The work of NR was partially supported by an A.P.Sloan
Foundation Fellowship and by NSF PYI DMR-9157484. The work of GM was supported
by DOE grant DE-AC02-76ER03075 and by a PYI.

\vspace*{\fill}
\end{document}